\documentclass{INTERSPEECH2023}

\usepackage{caption}
\usepackage{subcaption}
\usepackage{multirow}
\usepackage{graphicx}
\usepackage{cite}
\usepackage{nicefrac}

\interspeechcameraready

\title{End-to-End Zero-Shot Voice Conversion with \\ Location-Variable Convolutions}
\name{Wonjune Kang$^1$, Mark Hasegawa-Johnson$^2$, Deb Roy$^1$}
\address{
  $^1$Massachusetts Institute of Technology \quad $^2$University of Illinois at Urbana-Champaign
}
\email{}

\begin{document}
\newcommand\mdoubleplus{\ensuremath{\mathbin{+\mkern-10mu+}}}

\maketitle

\begin{abstract}
Zero-shot voice conversion is becoming an increasingly popular research topic, as it promises the ability to transform speech to sound like any speaker. However, relatively little work has been done on end-to-end methods for this task, which are appealing because they remove the need for a separate vocoder to generate audio from intermediate features. In this work, we propose LVC-VC, an end-to-end zero-shot voice conversion model that uses location-variable convolutions (LVCs) to jointly model the conversion and speech synthesis processes. LVC-VC utilizes carefully designed input features that have disentangled content and speaker information, and it uses a neural vocoder-like architecture that utilizes LVCs to efficiently combine them and perform voice conversion while directly synthesizing time domain audio. Experiments show that our model achieves especially well balanced performance between voice style transfer and speech intelligibility compared to several baselines.
\end{abstract}

\section{Introduction}

Voice conversion (VC) is the task of transforming a voice to sound like another person without changing the linguistic content in the original utterance~\cite{sisman2020overview}.
% It belongs to the general field of speech synthesis, which also includes text-to-speech (TTS), speech vocoding, and the changing of other speech properties such as emotion and accents.
It has many applications, such as voice anonymization, communication aids for the speech-impaired, and voice dubbing, which have contributed to its increasing popularity as a research direction in recent years.
Advances in deep learning have had a significant impact on voice conversion systems, allowing them to achieve major improvements in terms of voice quality and similarity to the target speaker, especially in the non-parallel data setting~\cite{hsu2017voice, kaneko2019cyclegan, van2020vector}.

Recently, attention has been shifting to zero-shot VC, a setting in which conversion is applied to new speakers that were previously unseen during training. A key aspect of zero-shot VC is the disentanglement, separation, and recombination of the content and speaker information in input and target speakers' utterances.
Most models are composed of content and speaker encoders that are trained to disentangle the two types of information, and a decoder that recombines them.
Traditionally, this decomposition and recombination are performed on intermediate representations such as mel spectrograms, followed by a vocoding step to generate time-domain audio. However, these approaches necessitate sequential training of the VC and vocoder stages using different supervision and criteria, leaving the potential for weaknesses in any individual module to cascade in the overall system.
Furthermore, they do not take advantage of data-driven end-to-end learning, which tends to bring simplicity and high performance. However, relatively little work has been done on end-to-end methods for voice conversion.

This work introduces an end-to-end approach for zero-shot voice conversion based on the architecture of a neural vocoder.
Our proposed model takes a set of carefully designed input features that have disentangled content and speaker information, and its vocoder-like architecture learns to combine them to perform voice conversion while simultaneously synthesizing audio. This significantly streamlines the model's structure and removes the difficult task of teaching it to perform disentangled representation learning.
Our model additionally utilizes location-variable convolutions (LVCs)~\cite{zeng2021lvcnet} as a core component, and we thus call it Location-Variable Convolution-based Voice Conversion (LVC-VC).
LVCs generate convolution kernels that are adaptive to the input conditional features and apply different convolution operations on different intervals of the input sequence.
This allows them to efficiently model the many time-dependent features that arise in speech using only a small number of parameters.
Experiments demonstrate that our model achieves competitive or better voice style transfer performance compared to several baselines while maintaining the clarity and intelligibility of transformed speech especially well.

The contributions of this paper are threefold: i) we propose a novel end-to-end model for zero-shot voice conversion based on the architecture of a neural vocoder, ii) we apply LVCs to the voice conversion task, showing that they enable efficient and interpretable combination of speaker and content information in the voice conversion process, and iii) we demonstrate that our model achieves a much better trade-off between audio quality and accurate voice style transfer compared to other baselines.\footnote{Code is available online at: \url{https://github.com/wonjune-kang/lvc-vc}. Audio samples are available on our demo page: \url{https://lvc-vc.github.io/lvc-vc-demo/}}

\vspace{-2pt}
\section{Background and Related Work}

\subsection{Zero-shot voice conversion}

\vspace{-1pt}

One of the first zero-shot VC models was AutoVC \cite{qian2019autovc}, an autoencoder-based model that utilizes a dimensionality bottleneck to disentangle content and speaker information. It has served as the base model for a range of extensions~\cite{qian2020f0, yuan2020improving, lee2021voicemixer}. Other approaches have used adaptive instance normalization~\cite{chou2019one} or activation function guidance~\cite{chen2021again} for information disentanglement. All of these approaches produce spectrograms and require a separate vocoder to synthesize time-domain audio.

Relatively few models have been proposed that can perform end-to-end voice conversion.
Blow \cite{serra2019blow} is a normalizing flow network for non-parallel raw-audio voice conversion. However, it is not able to perform zero-shot conversion, and like many other flow-based networks, it has a very large number of parameters. NVC-Net \cite{nguyen2021nvc} is a GAN-based zero-shot model that performs conversion directly on raw audio waveforms.
Conceptually, LVC-VC shares some similarities with HiFi-VC~\cite{kashkin2022hifi} and NANSY \cite{choi2021neural}. HiFi-VC is an end-to-end VC model that uses linguistic, F0, and speaker features as inputs to a conditioned HiFi-GAN \cite{kong2020hifi} decoder. NANSY is a non-end-to-end neural analysis and synthesis framework that uses various input features along with information perturbation-based training to control speech attributes. However, both methods use representations from large pre-trained ASR models for linguistic features (Conformer~\cite{gulati2020conformer}, 119M params, and XLSR-53 \cite{conneau2020unsupervised}, 317M params, respectively), resulting in very large model sizes. In contrast, LVC-VC has significantly fewer parameters and uses input features that are much simpler to compute.

\vspace{-2pt}
\subsection{Location-variable convolutions (LVCs)}
\vspace{-1pt}

Many speech generative models~\cite{oord2016wavenet, kumar2019melgan, kong2020hifi, jang2021univnet} are implemented using a WaveNet-like structure~\cite{oord2016wavenet}, in which dilated causal convolutions are applied to capture the long-term dependencies of a waveform. This necessitates a large number of convolution kernels to capture the many time-dependent features that arise in speech. However, in a traditional linear prediction vocoder \cite{atal1971speech}, the coefficients for the all-pole linear filter vary depending on the conditioning acoustic features of the analysis frame. A network with similarly variable kernels depending on the conditioning features could be able to model long-term dependencies in audio more efficiently than fixed-kernel methods.

Inspired by this idea, location-variable convolutions (LVCs)~\cite{zeng2021lvcnet} use different convolutional kernels to model different intervals in an input sequence depending on the corresponding ``local'' sections of a conditioning sequence. To do this, LVCs utilize kernel predictor networks which generate kernel weights given a conditioning sequence, such as a mel spectrogram. Then, each interval of the input sequence has a different convolution performed on it depending on the temporally associated section of the conditioning sequence. This gives LVCs more powerful capabilities for modeling long-term dependencies in audio because they can flexibly generate kernels that directly correspond to different conditioning sequences.
% For example, UnivNet \cite{jang2021univnet} demonstrated that LVCs could be used to achieve state-of-the-art performance in a GAN-based neural vocoder while using a relatively small number of parameters.

\vspace{-1pt}
\section{LVC-VC: Location-Variable Convolution-based Voice Conversion}
\vspace{-1pt}

We utilize a neural vocoder that incorporates LVCs~\cite{jang2021univnet} as the backbone architecture for LVC-VC. Taking appropriately designed content and speaker features as inputs to the LVC kernel predictors, our model efficiently combines their information to perform voice conversion while directly synthesizing audio.

\vspace{-4pt}
\subsection{Input features}
\label{ssec:features}
\vspace{-2pt}

\textbf{Content.}
The primary content feature used in LVC-VC is grounded in the source-filter model of speech production, in which the glottal excitation is convolved with the vocal tract impulse response~\cite{atal1971speech}. We treat the excitation as containing an utterance's speaker information and the vocal tract the content information.
Thus, we can largely disentangle the two by performing deconvolution to separate the source and filter components via low-quefrency liftering~\cite{childers1977cepstrum}.
Note that the filter will still contain some speaker information even after deconvolution; Section \ref{ssec:training} describes a strategy to deal with this.

Formally, let a source utterance in the time domain be $\mathbf{x}$ and its log-mel spectrogram be $\mathbf{X}$. We convert $\mathbf{X}$ to the cepstral domain, perform low-quefrency liftering, and re-convert back to the spectral domain to obtain the spectral envelope $\mathbf{H}$.
We also use the per-frame normalized quantized log F0, $\mathbf{p}_{\mathrm{norm}}$, from \cite{qian2020f0} as an additional content feature.

\vspace{2pt}

\noindent \textbf{Speaker.}
We use speaker embeddings extracted from a pre-trained encoder $E_s$; for an utterance $\mathbf{x}$, the speaker embedding is: $s = E_s(\mathbf{x})$. We describe $E_s$ in more depth in Section \ref{ssec:model}.
We also use a representation of a speaker's median log F0, $m$, quantized into 64 bins. Specifically, we quantize the range $\log 65.4$~Hz to $\log 523.3$~Hz (corresponding to the notes `C2' and `C5') and one-hot encode a speaker's median log F0; any values outside the quantized range are clipped.

\begin{figure}
    \centering
    \begin{subfigure}[b]{0.51\columnwidth}
        \centering
        \includegraphics[width=\textwidth]{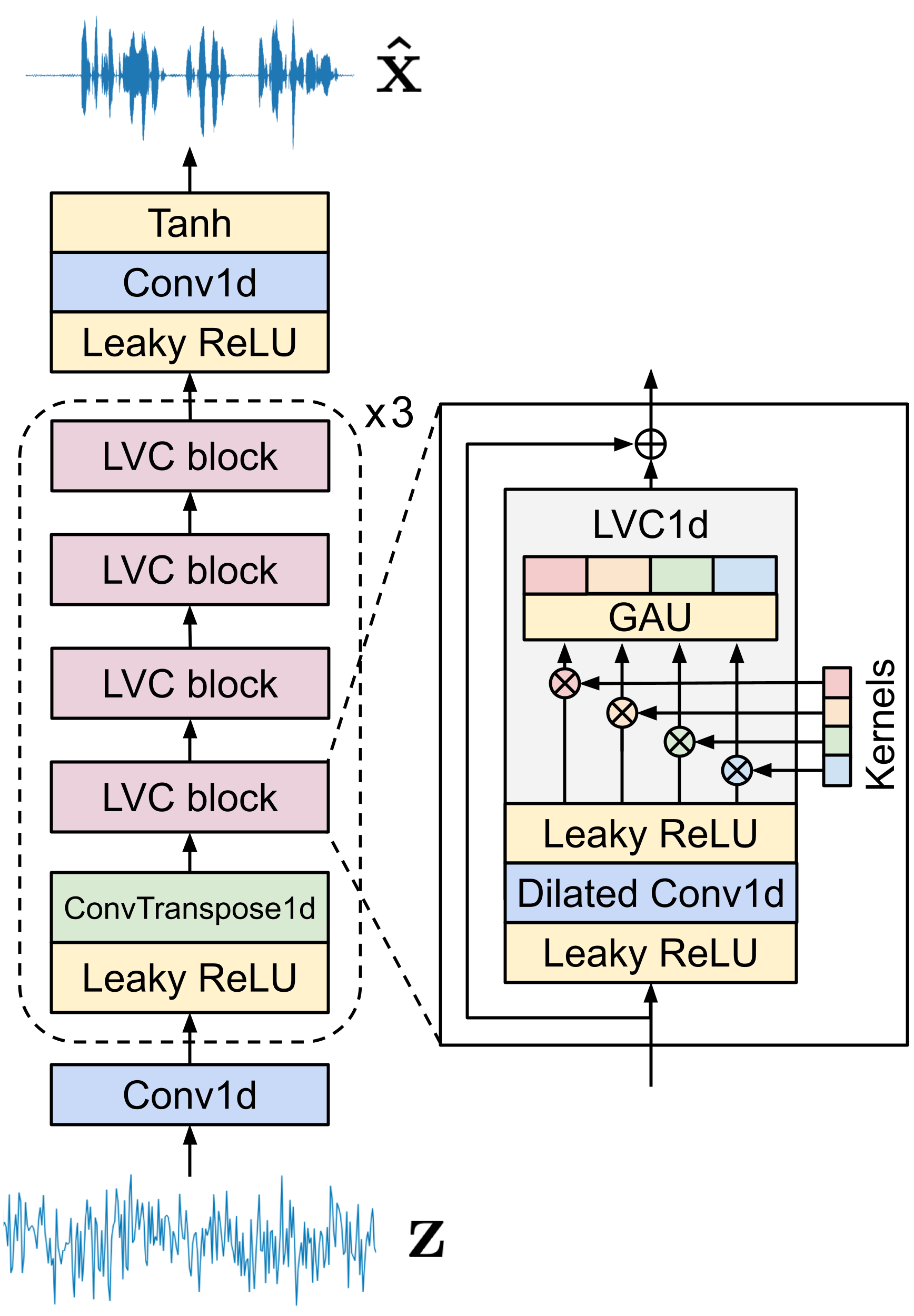}
        \caption{Generator.}
    \end{subfigure}
    \begin{subfigure}[b]{0.46\columnwidth}
        \centering
        \includegraphics[width=\textwidth]{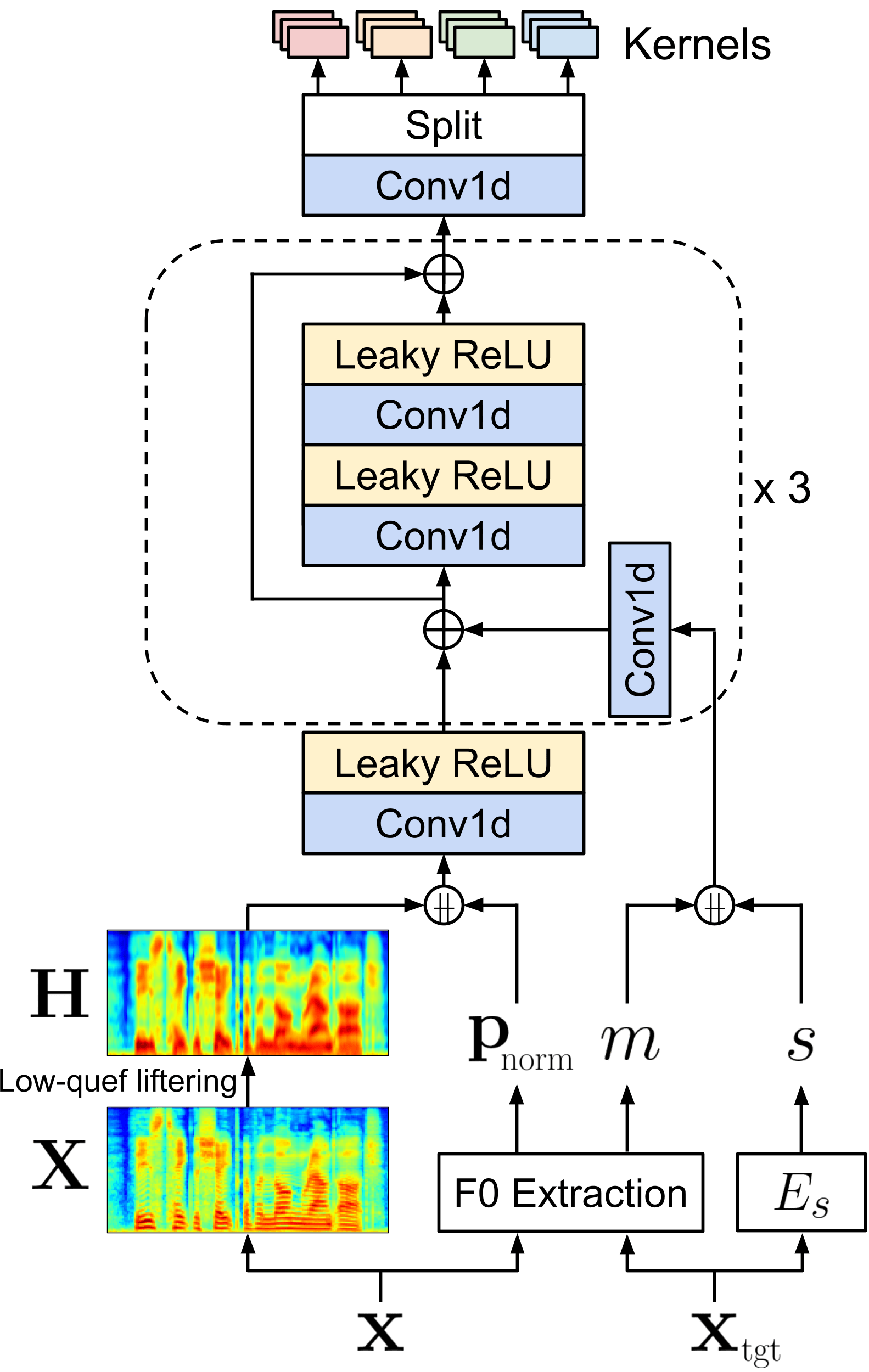}
        \caption{Kernel predictor for LVCs.}
    \end{subfigure}
    \vspace{-4pt}
    \caption{The components of the overall LVC-VC architecture. Content and speaker features are fed into the kernel predictors, which output kernels for the LVC layers in the generator. Each kernel predictor outputs the kernels for all four LVC blocks in a given transposed convolutional stack (shown in red, yellow, green, and blue at the right of (a) and top of (b)). $\mdoubleplus$ denotes stacking/concatenation in (b).}
    \label{fig:generator}
    \vspace{-12pt}
\end{figure}

\vspace{-4pt}
\subsection{Model architecture}
\label{ssec:model}
\vspace{-2pt}

LVC-VC consists of a generator $G$, a speaker encoder $E_s$, and a set of discriminators $D$ for GAN-based training.
% The input features are fed into the kernel predictor networks of the generator's LVC layers, which combine the information from the various features and pass it to the generator to synthesize audio.

\vspace{2pt}

\noindent \textbf{Generator.}
The generator $G$ is a fully convolutional neural network based on UnivNet-c16 \cite{jang2021univnet}.
Figure \ref{fig:generator} shows a diagram of its components. $G$ takes random noise $\mathbf{z}$ as an input sequence and the content and speaker features described in Section \ref{ssec:features} as conditions, and outputs a raw audio waveform $\mathbf{\hat{x}}$:
\vspace{-2pt}
\begin{equation*}
    \mathbf{\hat{x}} = G( \mathbf{z}, \mathbf{H}, \mathbf{p}_{\mathrm{norm}}, s, m ).
\end{equation*}
\vspace{-14pt}

\noindent It consists of three transposed convolutional stacks, each of which contains a 1D transposed convolution and four LVC blocks.
These serve to upsample $\mathbf{z}$, which is specified to have the same length as the content feature $\mathbf{H}$.
Kernel predictor networks take the conditioning features $\mathbf{H}$, $\mathbf{p}_{\mathrm{norm}}$, $s$, and $m$ as inputs and output the kernel weights for all of the LVC blocks in the stack that they are associated with.

\vspace{2pt}

\noindent \textbf{Speaker encoder.}
For the speaker encoder $E_s$, we use the Fast ResNet-34 speaker recognition model from \cite{chung2020defence}. The model was pre-trained on the development set of the VoxCeleb2 dataset \cite{chung2018voxceleb2} and uses self-attentive pooling to aggregate frame-level features into an utterance-level representation.

\vspace{2pt}

\noindent \textbf{Discriminators.}
We use a multi-resolution spectrogram discriminator (MRSD)~\cite{jang2021univnet} and a multi-period waveform discriminator (MPWD)~\cite{kong2020hifi} for GAN-based training.
The MRSD is a mixture of $M$ sub-discriminators that evaluates a synthesized waveform at multiple frequency resolution scales using various short-time Fourier transform (STFT) parameters.
The MPWD is also a mixture of sub-discriminators, each of which takes equally spaced samples of an audio waveform at a different period $p$ and evaluates whether it is real or not.
We use the same sets of STFT parameters and periods as in the original works.

\begin{table*}[t!]
    \centering
    \caption{Model sizes and voice conversion evaluation results on the three conversion settings. We include 95\% confidence intervals for MOS. Bold and underlined values indicate the best and second best scores in a given metric, respectively.}
    \vspace{-6pt}
    \fontsize{7.25}{8.25}\selectfont
    \begin{tabular}{lccccccccccccc}
    \toprule
    \multirow{3}{*}{\textbf{Model}} &
    \multirow{3}{*}{\textbf{\# Params}}  &
    \multicolumn{4}{c}{\textbf{Seen-to-Seen}}       &
    \multicolumn{4}{c}{\textbf{Unseen-to-Seen}}     &
    \multicolumn{4}{c}{\textbf{Unseen-to-Unseen}}   \\
    \cmidrule(lr){3-6}
    \cmidrule(lr){7-10}
    \cmidrule(lr){11-14}
    
    &
    &
    \textbf{MOS}      & \textbf{Sim}  & \textbf{CER}  & \textbf{EER}  &
    \textbf{MOS}      & \textbf{Sim}  & \textbf{CER}  & \textbf{EER}  &
    \textbf{MOS}      & \textbf{Sim}  & \textbf{CER}  & \textbf{EER}  \\ \midrule
    
    Ground Truth    &   --      & 4.61 {\scriptsize $\pm$ 0.11}             & 91.88             & 3.81              & 0.00              & 4.74 {\scriptsize $\pm$ 0.08}             & 95.63             & 2.93              & 0.00              & 4.62 {\scriptsize $\pm$ 0.11}             & 92.50                     & 3.58          & 0.00          \\
    UnivNet         &   --      & 4.33 {\scriptsize $\pm$ 0.12}             & 90.63             & 4.97              & 5.00              & 4.51 {\scriptsize $\pm$ 0.11}             & 97.50             & 3.59              & 0.00              & 4.58 {\scriptsize $\pm$ 0.10}             & 92.50                     & 4.68          & 0.00          \\ \midrule
    LVC-VC          &   5.97M   & \underline{3.54 {\scriptsize $\pm$ 0.17}} & 51.88             & \textbf{11.00}    & \underline{17.50} & \underline{3.51 {\scriptsize $\pm$ 0.18}} & 43.75             & \textbf{7.03}     & \underline{20.00} & 3.24 {\scriptsize $\pm$ 0.18}             & 38.75                     & \textbf{8.29}          & \textbf{26.25}         \\ \midrule
    AdaIN-VC        &   4.89M   & 2.35 {\scriptsize $\pm$ 0.16}             & \underline{63.75} & 22.78             & 28.75             & 2.57 {\scriptsize $\pm$ 0.17}             & \underline{53.13} & 16.41             & 36.25             & 2.41 {\scriptsize $\pm$ 0.16}             & \textbf{49.38}     & 20.60         & 35.00         \\
    AGAIN-VC        &   7.93M   & 2.13 {\scriptsize $\pm$ 0.16}             & 48.75             & 25.12             & 18.75             & 2.39 {\scriptsize $\pm$ 0.16}             & 46.25             & 23.94             & 31.25             & 2.26 {\scriptsize $\pm$ 0.15}             & \underline{42.50}  & 25.79         & \underline{31.25}         \\
    AutoVC          &   40.68M  & \textbf{3.84 {\scriptsize $\pm$ 0.15}}    & 30.63             & \underline{11.15} & 30.00             & \textbf{3.71 {\scriptsize $\pm$ 0.16}}    & 31.88             & 10.65             & 26.25             & \textbf{3.61 {\scriptsize $\pm$ 0.17}}    & 13.13                     & \underline{12.07} & 63.75         \\
    AutoVC-F0       &   41.21M  & 3.44 {\scriptsize $\pm$ 0.16}             & 32.50             & 12.53             & 28.75             & 3.39 {\scriptsize $\pm$ 0.16}             & 37.50             & \underline{10.54} & 32.50             & \underline{3.31 {\scriptsize $\pm$ 0.17}} & 20.00                     & 14.20         & 65.00         \\
    Blow            &   62.11M  & 1.78 {\scriptsize $\pm$ 0.15}             & 29.38             & 18.33             & 52.50             & --                                        & --                & --                & --                & --                                        & --                        & --            & --            \\
    NVC-Net         &   15.13M  & 2.96 {\scriptsize $\pm$ 0.19}             & \textbf{76.88}    & 31.46             & \textbf{12.50}    & 3.14 {\scriptsize $\pm$ 0.19}             & \textbf{66.25}    & 26.91             & \textbf{11.25}    & 3.10 {\scriptsize $\pm$ 0.20}             & 40.00                     & 26.27         & 37.50         \\
    \bottomrule
    \end{tabular}
    \label{table:results}
    \vspace{-10pt}
\end{table*}

\vspace{-2pt}
\subsection{Training}
\label{ssec:training}
\vspace{-1pt}

\noindent \textbf{Self-reconstruction.}
LVC-VC is trained primarily using self-reconstruction, where the content and speaker features from training utterances are used to reconstruct the original audio.
Recall that the content feature $\mathbf{H}$ still contains some residual speaker information, which could leak into the synthesized audio at inference time. To prevent this, we warp $\mathbf{H}$ by stretching or compressing it along the frequency axis during training; we denote the warped version $\mathbf{H}'$. We found that this removes most of the residual speaker information in $\mathbf{H}$ while still preserving its content information.
The warping factor is randomly chosen from a uniform distribution over $[0.85, 1.15]$ for each sample.
For speaker embeddings, rather than extracting them directly from an utterance, we sample an embedding $s'$ from a 1-component Gaussian that we fit on the corresponding speaker's training utterances. This means that a similar, but different embedding is used to reconstruct an utterance every time, which we found to help with generalization to unseen speakers.

Let a training utterance be $\mathbf{x}$ and its associated features be $\mathbf{H}', \mathbf{p}_{\mathrm{norm}}, s', m$. Then, the reconstructed output is $\mathbf{\hat{x}} = G( \mathbf{z}, \mathbf{H}', \mathbf{p}_{\mathrm{norm}}, s', m )$. In addition to the discriminator losses, we use multi-resolution STFT loss \cite{yamamoto2020parallel}. The full auxiliary reconstruction loss $\mathcal{L}_\mathrm{aux}$, which consists of spectral convergence loss $\mathcal{L}_\mathrm{sc}$ and log STFT magnitude loss $\mathcal{L}_\mathrm{mag}$, is:
\begin{align*}
    \mathcal{L}_{\mathrm{aux}}(\mathbf{x}, \mathbf{\hat{x}}) = \frac{1}{M} \sum_{m=1}^{M} \mathbb{E}_{\mathbf{x}, \mathbf{\hat{x}}} \Big[ \mathcal{L}_\mathrm{sc}( \mathbf{s}_m, \mathbf{\hat{s}}_m ) + \mathcal{L}_\mathrm{mag}( \mathbf{s}_m, \mathbf{\hat{s}}_m ) \Big], \\
    \mathcal{L}_\mathrm{sc}( \mathbf{s}, \mathbf{\hat{s}} ) = \frac{\lVert \mathbf{s} - \mathbf{\hat{s}} \rVert_F}{\lVert \mathbf{s} \rVert_F}, \quad
    \mathcal{L}_\mathrm{mag}( \mathbf{s}, \mathbf{\hat{s}} ) = \frac{1}{N} \lVert \log \mathbf{s} - \log \mathbf{\hat{s}} \rVert_1.
\end{align*}
Here, $N$ denotes the number of spectrogram frames and
% $\norm \cdot \norm_F$ and $\norm \cdot \norm_1$ denote the Frobenius and L1 norms, respectively.
$M$ is the number of MRSD sub-discriminators. $\mathbf{s}$ and $\mathbf{\hat{s}}$ are the ground truth and predicted spectrograms in the MRSD, respectively, and each $m$-th $\mathcal{L}_\mathrm{sc}$ and $\mathcal{L}_\mathrm{mag}$ reuse $\mathbf{s}_m$ and $\mathbf{\hat{s}}_m$ from the $m$-th MRSD sub-discriminator.

\vspace{2pt}

\noindent \textbf{Speaker similarity.}
We use a speaker similarity criterion (SSC) in order to induce generation of audio that more closely matches the characteristics of the target speaker.
Let a given utterance for self-reconstruction be $\mathbf{x}_0$ and its associated features be $(\mathbf{H}_0, \mathbf{p}_{\mathrm{norm, 0}}, s_0, m_0)$. We sample $N$ utterances from different speakers $\mathbf{x}_1, ..., \mathbf{x}_N$ with content features $(\mathbf{H}_n, \mathbf{p}_{\mathrm{norm}, n})$, $\forall n \in [1, ..., N]$.
Then, the SSC loss $\mathcal{L}_{\mathrm{ssc}}$ is:
\begin{align*}
    \mathcal{L}_{\mathrm{ssc}} &= \frac{1}{N} \sum_{n=1}^{N} \cos \big( E_s(\mathbf{\hat{x}}_{n \rightarrow 0}), s'_0 \big), \\
    \mathbf{\hat{x}}_{n \rightarrow 0} &= G( \mathbf{z}, \mathbf{H}_n, \mathbf{p}_{\mathrm{norm}, n}, s'_0, m_0 ),
    % \mathcal{L}_{\mathrm{ssc},-} &= -\frac{1}{N} \sum_{n=0}^{N-1} \cos \big( E_s(\mathbf{\hat{x}}_{n,N}), s_n \big),
\end{align*}
where $\cos(x_1, x_2)$ is the cosine similarity between $x_1$ and $x_2$.

\vspace{2pt}

\noindent \textbf{Overall criteria.}
During training, we keep the weights of $E_s$ fixed and only update $G$ and $D$. The overall generator and discriminator losses follow the LSGAN objective functions \cite{mao2017least}:
\begin{align*}
    \mathcal{L}_{G} = &\frac{1}{K} \sum_{k=1}^{K} \mathbb{E}_{\mathbf{z}, \mathbf{c}}\left[\left(D_{k}(G(\mathbf{x}'))-1\right)^{2}\right] \nonumber \\
    &+ \lambda_{\mathrm{aux}} \mathcal{L}_{\mathrm{aux}}(\mathbf{x}, G(\mathbf{x}')) + \lambda_{\mathrm{ssc}} \mathcal{L}_{\mathrm{ssc}}, \\
    \mathcal{L}_{D} = &\frac{1}{K} \sum_{k=1}^{K}\left(\mathbb{E}_{\mathbf{x}}\left[\left(D_{k}(\mathbf{x})-1\right)^{2}\right] + \mathbb{E}_{\mathbf{z}, \mathbf{c}}\left[D_{k}(G(\mathbf{x}'))^{2}\right]\right),
\end{align*}
where we abbreviate $G( \mathbf{z}, \mathbf{H}', \mathbf{p}_{\mathrm{norm}}, s', m )$ to $G(\mathbf{x}')$ for conciseness. $K$ is the number of all sub-discriminators across the MRSD and MPWD, and $D_k$ denotes the $k$-th sub-discriminator. $\lambda_{\mathrm{aux}}$ and $\lambda_{\mathrm{ssc}}$ are weighting factors that balance the contributions of the auxiliary loss and SSC loss, respectively.

\subsection{Inference}

Given source utterance $\mathbf{x}_\mathrm{src}$ with content features $(\mathbf{H}_\mathrm{src},$ $\mathbf{p}_{\mathrm{norm, src}})$ and target utterance $\mathbf{x}_\mathrm{tgt}$ with speaker features $(s_\mathrm{tgt},$ $m_\mathrm{tgt})$, the converted utterance $\mathbf{\hat{x}}_{\mathrm{src} \rightarrow \mathrm{tgt}}$ is produced by:
\begin{equation*}
    \mathbf{\hat{x}}_{\mathrm{src} \rightarrow \mathrm{tgt}} = G( \mathbf{z}, \mathbf{H}_\mathrm{src}, \mathbf{p}_{\mathrm{norm, src}}, s_\mathrm{tgt}, m_\mathrm{tgt} ).
\end{equation*}
\section{Experiments}

\subsection{Configurations}
\label{ssec:config}

\textbf{Data.} We used the VCTK Corpus \cite{yamagishi2019vctk} for training and evaluation.
% , which consists of around 44 hours of speech from 109 speakers.
% Of the 109 speakers, 47 are male and 62 are female.
All audio was resampled to 16 kHz. We randomly partitioned the data into 99 seen and 10 unseen speakers, and the seen speakers' utterances were further split into train and test sets in a 9:1 ratio. Only the seen speakers' train set was used for training.
We computed 80-dim log-mel spectrograms using a 1024 point Fourier transform, with a Hann window of size 1024 and hop length of size 256.
To obtain $\mathbf{H}$, we took the 20 lowest quefrency coefficients for low-quefrency liftering.

\vspace{2pt}

\noindent \textbf{Training.} All models were trained on a single NVIDIA RTX 3090 Ti GPU. We used the AdamW optimizer~\cite{loshchilov2018decoupled} with learning rate 1e-4 and $\beta_1 = 0.5, \beta_2 = 0.9$.
% All input features were cropped or padded to correspond to 16,384 samples for batch processing.
For the SSC loss, we set $N = 8$. Following \cite{jang2021univnet}, we set $\lambda_{\mathrm{aux}} = 2.5$, and we empirically set $\lambda_{\mathrm{ssc}} = 0.9$.
We trained our model with only self-reconstructive loss using a batch size of 32 for 1.8M iterations. Then, we halved the learning rate to 5e-5 and continued training with the SSC loss for 5,000 iterations.
This ensured that the model first learned to produce high-quality audio before being guided to perform better voice conversion without compromising audio quality. $\lambda_{\mathrm{ssc}}$ was linearly annealed from 0 to its final value for the first 2,000 steps in which the SSC loss was used.

\vspace{2pt}

\noindent \textbf{Evaluation.} We conducted subjective listening tests for naturalness and speaker similarity on Amazon MTurk; a total of 97 individuals participated. For naturalness, subjects provided a mean opinion score (MOS) on a scale from 1 to 5. For similarity, we used the binary ``same/different'' metric from \cite{wester2016analysis}. We also evaluated character error rate (CER) on automatic speech recognition (ASR) and equal error rate (EER) on automatic speaker verification (ASV).
For ASR, we used a pre-trained wav2vec 2.0 \textsc{Base} model \cite{baevski2020wav2vec} from the Hugging Face Transformers library \cite{wolf2020transformers}.
For ASV, we used a ResNet-34-based model~\cite{heo2020clova} that was trained on VoxCeleb2~\cite{chung2018voxceleb2}.

We considered three source-to-target conversion settings: seen-to-seen (s2s), unseen-to-seen (u2s), and unseen-to-unseen (u2u). For each setting, we evaluated conversions from 80 utterance pairs; each conversion was rated by two annotators.
% we sampled 10 source speakers and randomly assigned one speaker from each gender to be a target speaker. Then, for each source-target pair, we randomly sampled two utterances from each speaker to perform conversion, resulting in 80 (= 10 $\times$ 2 $\times$ 4) utterance pairs.

\begin{table}[t!]
    \centering
    \caption{Unseen-to-unseen conversion results for ablations.}
    \vspace{-6pt}
    \fontsize{7.6}{8.7}\selectfont
    \begin{tabular}{lccccccc}
    \toprule
    \multicolumn{1}{l}{\textbf{Model}}  & \textbf{CER}  & \textbf{EER}  & \textbf{NISQA}    \\ \midrule
    LVC-VC                              & 8.29          & 26.25         & 3.50 {\scriptsize $\pm$ 0.13}   \\ \midrule
    w/o Gaussian embeddings                  & 11.11         & 25.00         & 2.89 {\scriptsize $\pm$ 0.14}   \\
    w/o SSC loss                        & 6.64          & 68.75         & 3.83 {\scriptsize $\pm$ 0.13}   \\
    w/o warping $\mathbf{H}$            & 7.39          & 51.25         & 3.62 {\scriptsize $\pm$ 0.17}   \\
    w/o $\mathbf{p}_{\mathrm{norm}}$    & 8.60          & 32.50         & 3.36 {\scriptsize $\pm$ 0.18}   \\
    w/o $m$                             & 9.90          & 28.75         & 3.47 {\scriptsize $\pm$ 0.14}   \\
    \bottomrule
    \end{tabular}
    \label{table:ablation_u2u}
    \vspace{-10pt}
\end{table}

\subsection{Results}
\label{ssec:results}

We compared LVC-VC against AdaIN-VC~\cite{chou2019one}, AGAIN-VC~\cite{chen2021again}, AutoVC~\cite{qian2019autovc}, AutoVC-F0~\cite{qian2020f0}, Blow~\cite{serra2019blow}, and NVC-Net~\cite{nguyen2021nvc}.
For a fair comparison, all baselines were trained from scratch using the same data and spectrogram configurations as our model. For models requiring a vocoder, time-domain audio was synthesized using a UnivNet-c16 vocoder~\cite{jang2021univnet} that was trained on the 99 seen speakers from the VCTK dataset and the \textit{train-clean-360} split of the LibriTTS dataset~\cite{zen2019libritts}.

Table \ref{table:results} shows the scores for all models on the three conversion settings.
% We also report scores for ground truth speech and speech reconstructed using the UnivNet vocoder to provide reference values.
We found that most of the baselines either perform voice style transfer (VST) well but produce poor quality audio (AdaIN-VC, AGAIN-VC, NVC-Net), or produce high quality audio but do not perform VST well (AutoVC, AutoVC-F0).
In other words, there is a trade-off between audio quality and VST performance.
LVC-VC manages this trade-off much better than the other models. 
It achieves competitive performance with the other best models in terms of MOS and Similarity while consistently obtaining the lowest CER, especially in the u2s and u2u settings. This demonstrates that it maintains intelligibility especially well.
Notably, it arguably achieves the most balanced performance overall, even with a compact model size. 

\vspace{-4pt}
\subsection{Ablation studies}
\vspace{-2pt}

We conducted ablation studies on various aspects of LVC-VC; the results are shown in Table \ref{table:ablation_u2u}. For brevity and convenience, we only report scores from objective metrics in the u2u setting. As a proxy measure for subjective MOS, we used NISQA \cite{mittag2021nisqa}, which provides an estimate of an utterance's speech quality on a scale from 1 to 5.
We found that each of the ablated components contributed meaningfully to the model's performance. Training on fixed speaker embeddings instead of sampling from a Gaussian caused audio quality to degrade, suggesting that training on more diverse embeddings helps the model generalize better to new speakers. Training without the SSC loss or without warping $\mathbf{H}$ caused VST performance to decrease.
% showing the importance of explicitly guiding the model to perform conversion rather than only relying on self-reconstructive training, as well as perturbing the source speaker information in $\mathbf{H}$.
% so that the model does not learn to use the residual speaker information in that feature.
Finally, $\mathbf{p}_{\mathrm{norm}}$ and $m$ contributed to general performance gains in all metrics.

\vspace{-4pt}
\subsection{Analysis of the speech synthesis process}
\label{ssec:synthesis_analysis}
\vspace{-1pt}

\begin{figure}[t]
    \centering
    \begin{subfigure}{0.3\columnwidth}
        \centering
        \includegraphics[height=0.8in]{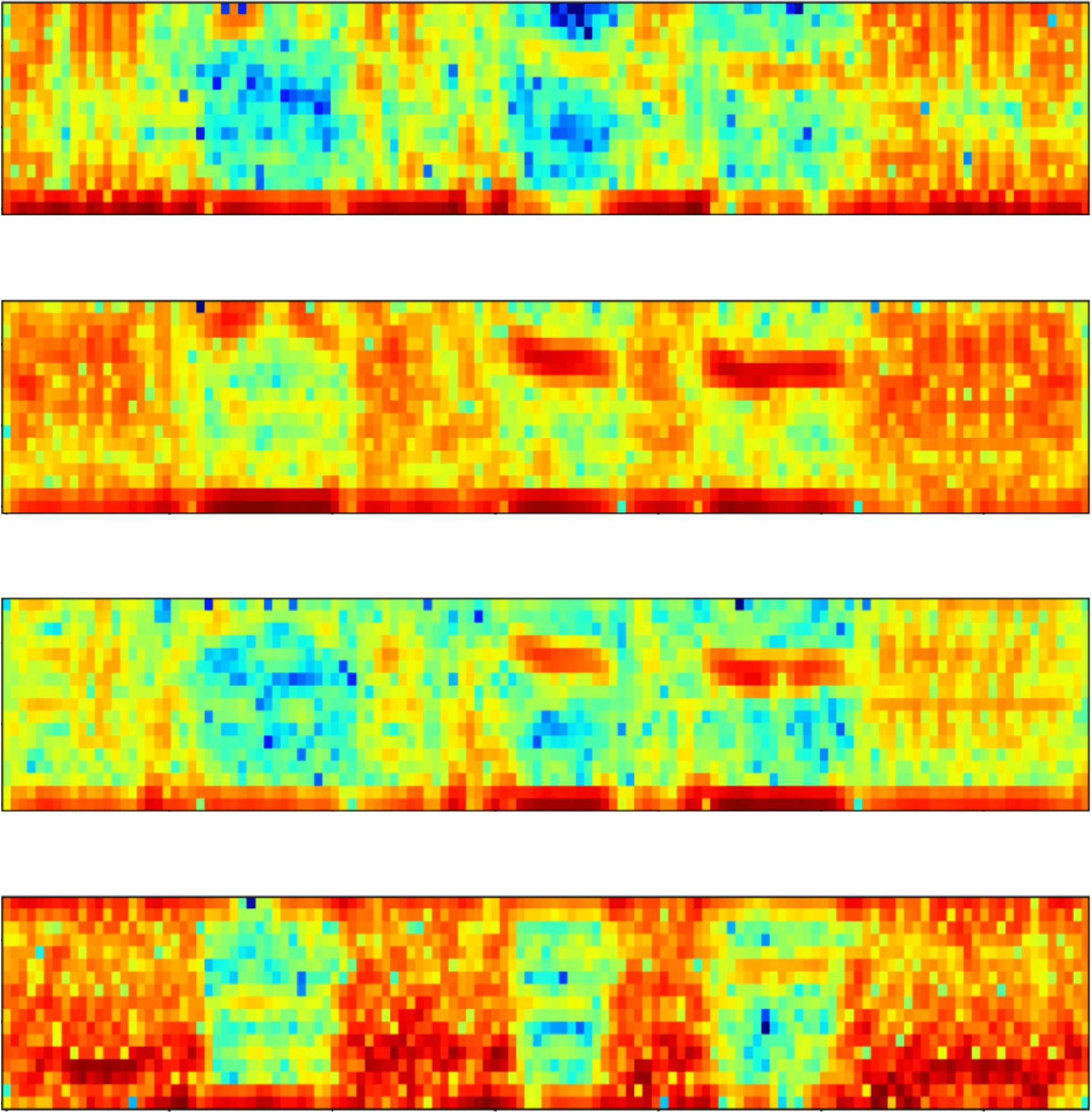}
        \caption{}
    \end{subfigure}
    \begin{subfigure}{0.34\columnwidth}
        \centering
        \includegraphics[width=\textwidth]{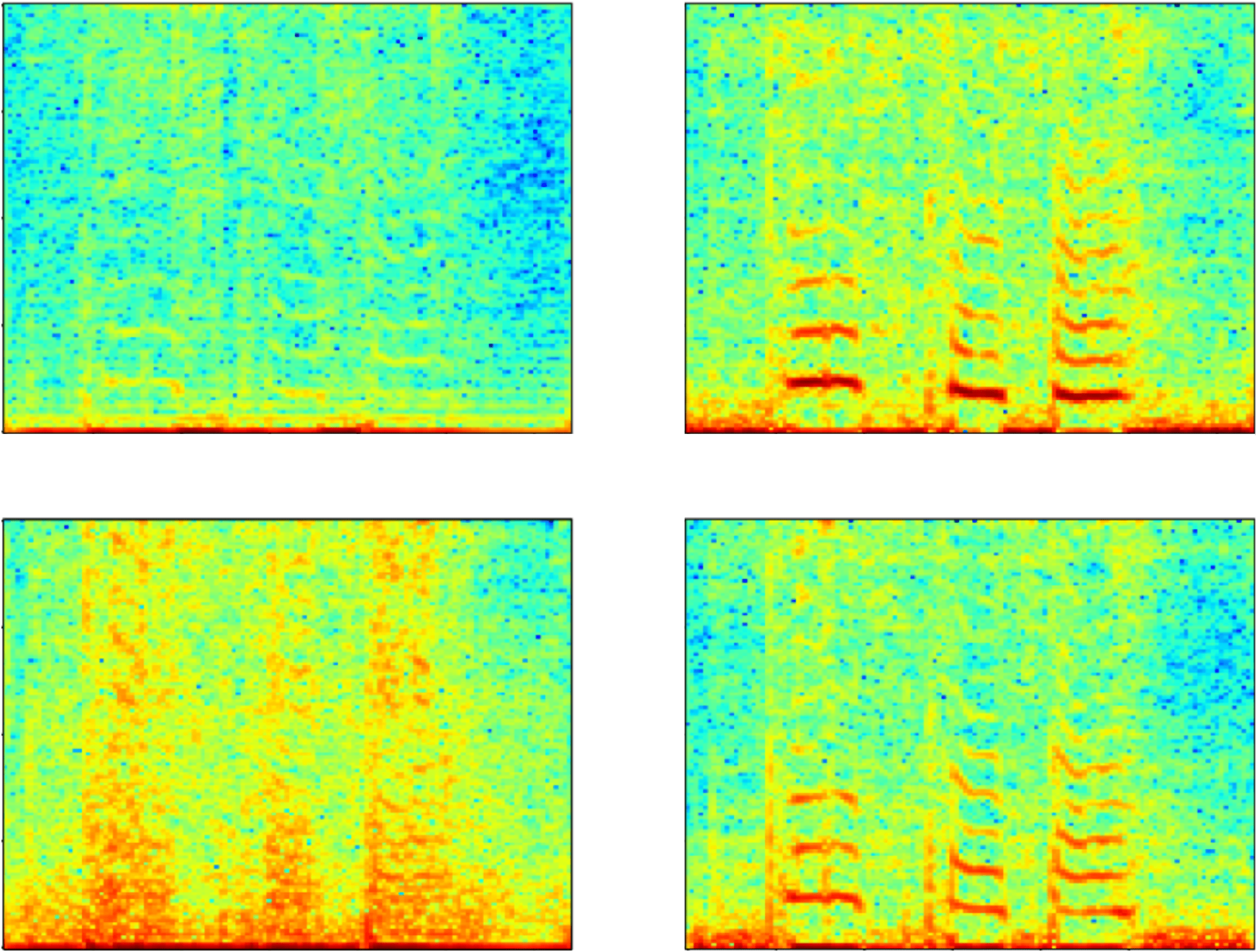}
        \caption{}
    \end{subfigure}
    \begin{subfigure}{0.33\columnwidth}
        \centering
        \includegraphics[width=\textwidth]{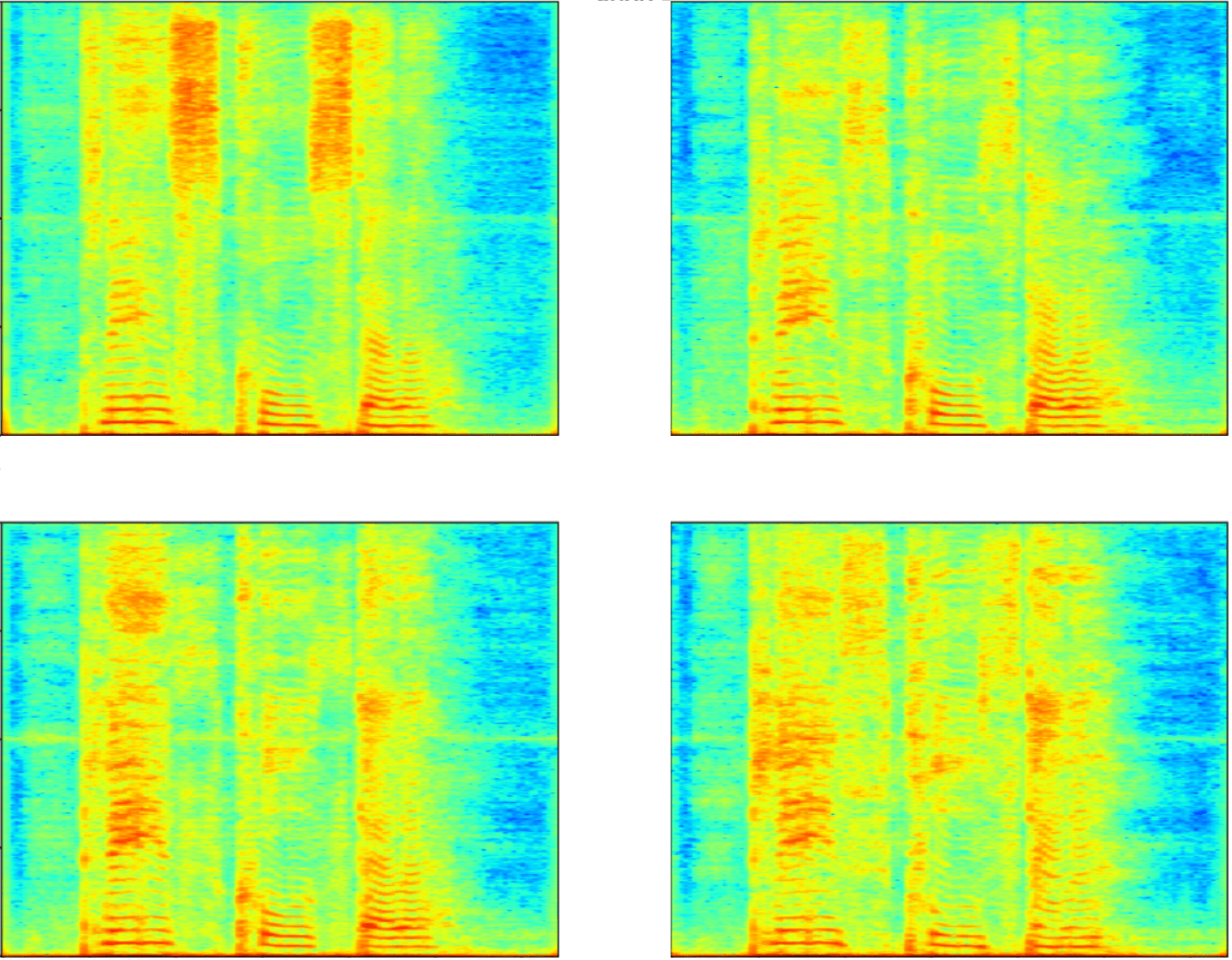}
        \caption{}
    \end{subfigure}
    \vspace{-5pt}
    \caption{Results of computing STFTs on outputs of the (a) first, (b) second, and (c) third transposed convolutional stacks. For brevity, we show only 4 of the 16 channels.}
    \label{fig:intermediate_stft}
    \vspace{-6pt}
\end{figure}

We investigated how LVC-VC generates audio by performing spectral analyses of its intermediate outputs. Specifically, we made the model reconstruct an utterance and performed STFTs on the outputs of each channel after each transposed convolutional stack. Figure \ref{fig:intermediate_stft} shows the results. In the first stack, we see that the model begins to incorporate content and speaker information from the emergence of voiced segments and the F0 band, respectively. Similar patterns emerge in subsequent stacks, with the formation of the F0 contour and more harmonics. This suggests the gradual addition of more detailed speaker and content information as the signal is upsampled. Individual channels also appear to model different aspects of the signal, such as voiced and unvoiced segments, formants, and background noise.

\begin{figure}[t]
    \centering
    \includegraphics[width=0.32\columnwidth]{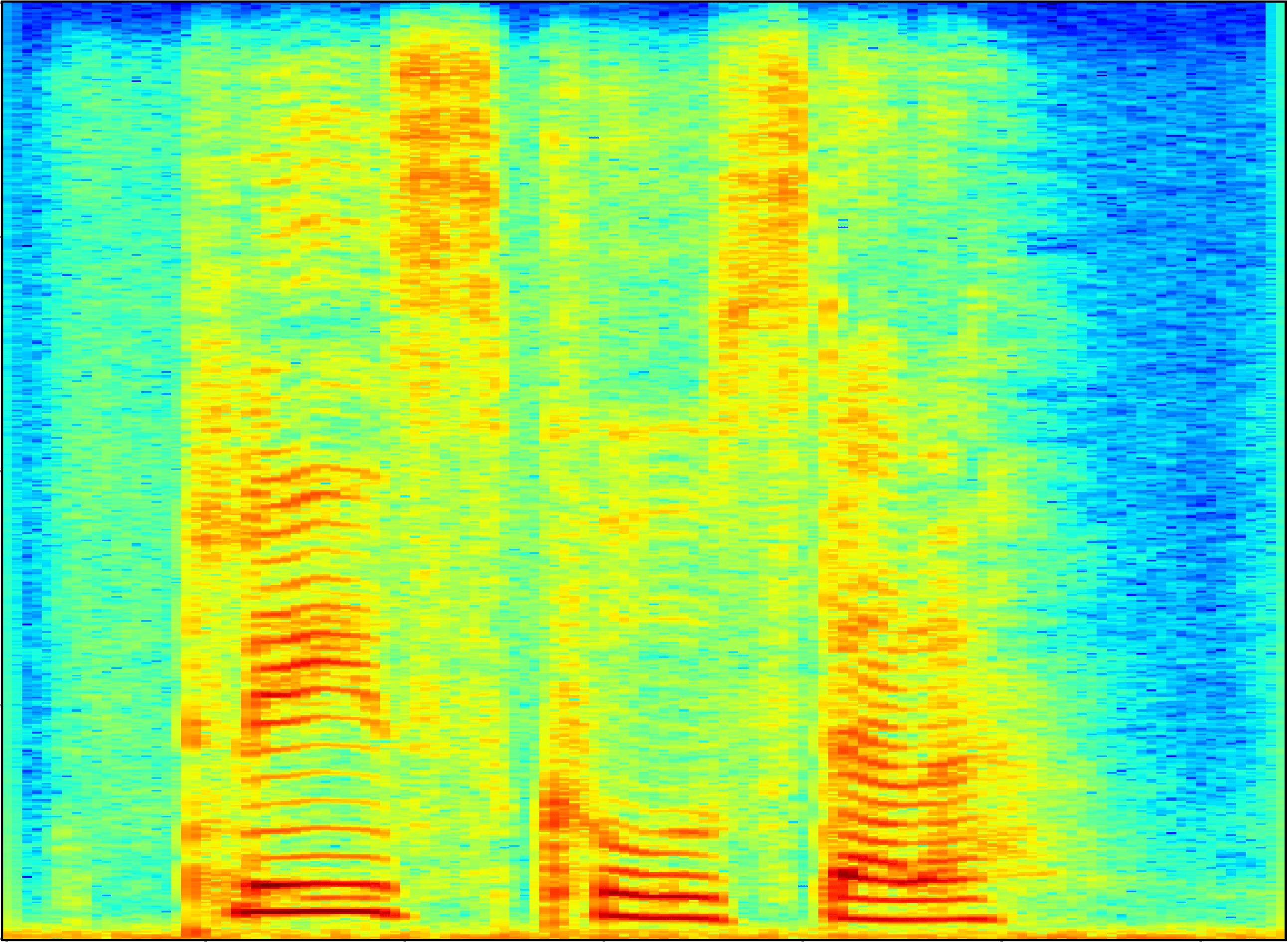}
    \hspace{0.1pt}
    \includegraphics[width=0.32\columnwidth]{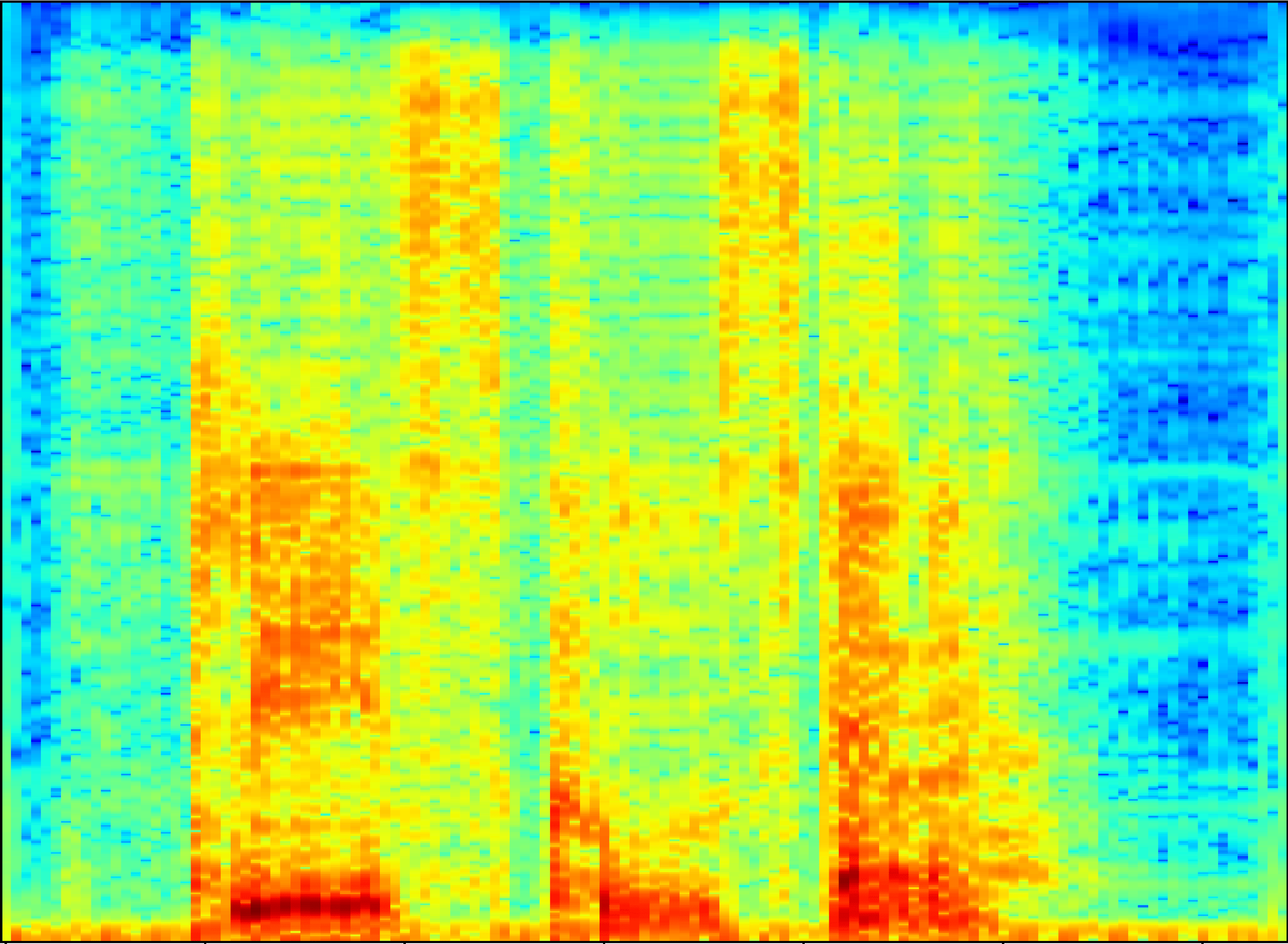}
    \hspace{0.1pt}
    \includegraphics[width=0.32\columnwidth]{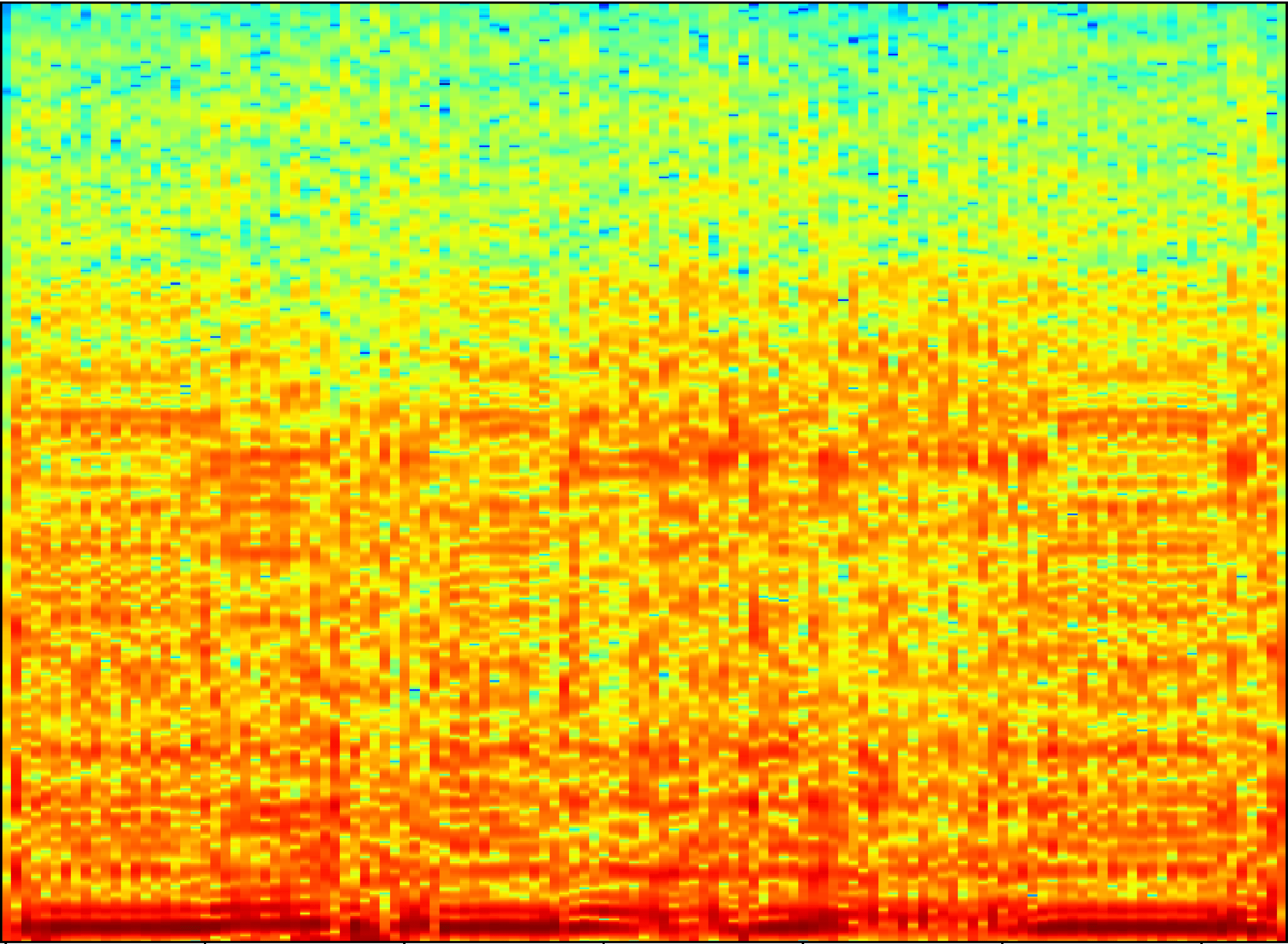}
    \vspace{-14pt}
    \caption{From left to right: spectrograms of the original utterance, audio generated when zeroing out speaker features, and audio generated when zeroing out content features.}
    \label{fig:lvc_zeroed_outputs}
    \vspace{-14pt}
\end{figure}

We also investigated how LVC-VC utilizes speaker and content information by making the model generate speech with either the speaker or content features zeroed out. Figure \ref{fig:lvc_zeroed_outputs} shows spectrograms of the resulting audio. As we might expect, when speaker information is zeroed out, the spectral envelope and formants appear, but the F0 and harmonics do not. Conversely, zeroing out content information leads to the F0 and harmonics still somewhat forming, but the spectral envelope and formants do not.
These results indicate that LVC-VC learns to independently apply speaker and content information towards generating time-domain audio in an interpretable way.

\vspace{-2.5pt}
\section{Conclusion}
\label{sec:conclusion}
\vspace{-1.5pt}

In this work, we presented LVC-VC, an end-to-end model for zero-shot voice conversion. LVC-VC utilizes a set of carefully designed input features that have disentangled content and speaker style information. Using location-variable convolutions, it combines this information within a neural vocoder-like framework, simultaneously performing voice conversion while generating audio. Despite having a compact model size, LVC-VC achieves competitive or better performance compared to several baselines and demonstrates the ability to maintain intelligibility especially well, thus achieving the most balanced performance overall. This demonstrates the effectiveness of the model at utilizing and combining the relevant information in the input features for speech synthesis.

% \section{Acknowledgements}

% \ifinterspeechfinal
%      The INTERSPEECH 2023 organisers
% \else
%      The authors
% \fi
% would like to thank ISCA and the organising committees of past INTERSPEECH conferences for their help and for kindly providing the previous version of this template.

\newpage
\bibliographystyle{IEEEtran}
\bibliography{bibliography.bib}

\end{document}